\documentclass[12pt]{iopart}
\usepackage{graphicx,epsf}
\usepackage{iopams}

\input{psfig}

\newcommand{\p}{\prime}
\newcommand{\para}{\parallel}

\newcommand{\beq}{\begin{equation}}
\newcommand{\eeq}{\end{equation}}
\newcommand{\bea}{\begin{eqnarray}}
\newcommand{\eea}{\end{eqnarray}}

\begin{document}

\title[Current density shaping by turbulence]{Turbulent Contributions to Ohm's Law in Axisymmetric Magnetized
Plasmas}

\author{I Chavdarovski$^{1}$, R Gatto$^{2}$}

\address{$^1$ {\it Max Planck Institute for Plasma Physics, Boltzmannstrasse 2,
Garching, 85748, Germany}}
\address{$^2$ {\it Department of Astronautical, Electrical and Energy Engineering, Sapienza University of Rome, Italy}}

\vspace{0.5cm}

\address{E-mail: chavdarovski@gmail.com, renato.gatto@uniroma1.it}

\begin{abstract}

The effect of magnetic turbulence in shaping the current density in
axisymmetric magnetized plasma
is analyzed using a turbulent extension of Ohm's law derived from the self-consistent
action-angle transport theory. Besides the well-known hyper-resistive (helicity-conserving)
contribution, the generalized Ohm's law contains an anomalous resistivity term, 
and a turbulent bootstrap-like term proportional to the current density derivative. 
The numerical solution of the
equation for equilibrium and
turbulence profiles characteristic of conventional and advanced scenarios shows that,
trough ``turbulent bootstrap'' effect and anomalous resistivity turbulence can generate power 
and parallel current which are a sizable portion
(about $20-25\, \%$) of the
corresponding effects associated with the neoclassical bootstrap effect.
 The degree of alignment of the turbulence peak and the pressure gradient plays an important
role in defining the steady-state regime. In fully bootstrapped tokamak,
the hyper-resistivity is essential in overcoming the intrinsic limitation of the hollow current profile.

\vspace{0.25cm}
\noindent
{\bf PACS Numbers:}
\end{abstract}

\vspace{0.5cm}

\section{Introduction}

International research efforts on achieving the necessary
conditions for controlled thermonuclear fusion in magnetically confined
plasmas have shown that in order for a steady-
state tokamak reactor to be economically attractive
high fraction of the current needs to be generated
non-inductively. The low drive efficiencies
of the known external non-inductive methods, such as
radio-frequency-wave or neutral beams injection, raise the interest in regimes
in which the current is generated internally via the bootstrap mechanism.
The bootstrap current is a parallel current driven by the radial
pressure gradient through the pressure anisotropy in toroidal geometry \cite{Galeev68}.
In ideal case, a high pressure tokamak would be able to generate all of its
current via the bootstrap effect. Obvious setback to this scenario is the fact
that the bootstrap current vanishes around the axis,
where the pressure profile is flat and trapped particle
population small \cite{Bickerton71}. Since a tokamak with large
bootstrap current is ordinarily unstable to tearing modes,
it has been suggested that poloidal flux generated spontaneously near
the edge by the dynamo effect induced by the turbulent perturbations,
can overcome this intrinsic limitation by diffusing the bootstrap
current toward the center.
This idea has been the driving force behind many theoretical
works \cite{Weening92,Itoh93}, as well as experimental campaigns aimed at
achieving, via optimization
of the plasma profiles, a steady-state ``fully bootstrapped tokamak'' operation
 \cite{Turnbull98, Sakamoto05,Coda08}.

The role of turbulence in explaining
the anomalous current diffusion observed in
experiments, is not yet clear.  It has been shown
in various theoretical works that turbulence leads to three additional terms in the Ohm's law:
(i) hyper-resistivity, a viscous-like term which induces current diffusion towards the axis \cite{Schmidt71},
(ii) anomalous resistivity which adds to the
neoclassical resistivity \cite{Biskamp84}, and
(iii) a bootstrap-like term \cite{GattoOPPJ2011}
which generates current density due to the transfer of linear momentum to the electrons
at the expense of the energy in the turbulent perturbations \cite{Nunan94,Fowler95}.
The scope of this work is to identify
scenarios in which the turbulent contributions play
significant role in regularizing the current density profile and sustaining
the equilibrium.

In this work we present numerical studies based on
a turbulent extension of Ohm's law \cite{GattoOPPJ2011}
derived in the framework of
the action-angle \cite{Kaufman72} self-consistent \cite{Mynick89} transport
theory. The self-consistency implies the collision operator contains both
diffusion and drag in action-space, as opposed to the quasi-linear
approach which includes only the diffusion part.
The radial structure of the turbulent transport
coefficients is presented, as a function of the magnetic turbulence
and the thermodynamic equilibrium profiles.
The hyper-resistive term is only related to the momentum transport,
while the anomalous and cross-resistivity (bootstrap-like) coefficients contain also terms originating
from the electron momentum source. The cross-resistive term leads to an
amplification of the total current,
while the anomalous and hyper-resistivity to a current reduction.
However, we show that the anomalous resistivity can
increase significantly the current in the outer region and in some scenarios, like the advanced,
even generate power from turbulence instead of dissipating it.

The present work continues the research line of Ref.~\cite{GattoOPPJ2011} extending it to
include advanced plasma regimes and fully bootstrapped tokamaks.
In Sec.~\ref{theory} we give a very brief overview of the theoretical transport
model and then present the turbulent electric field, referring the
interested reader to Ref.~\cite{GattoOPPJ2011} for the detailed
derivation of it. The significance of the various turbulent contributions is
discussed in Sec.~\ref{coeff} where we present their radial dependence for different
regimes and turbulence profiles. In Sec.~\ref{balance} we give the final form of the
turbulent Ohm's law and describe a power balance equation derived from it.
A series of numerical studies that consider various thermodynamic
profiles characteristic of L-mode regimes, advanced scenarios and fully bootstrapped
tokamak follows in Sec.~\ref{numerical}. These studies clarify the potential role that
the effect of turbulence
has in shaping current density and safety factor profiles and show how turbulence can
provide explanation to some experimental observations.
The summary and the conclusions are presented in Sec.~\ref{summary}.

\section{Theoretical model}
\label{theory}

The parallel momentum transport equation stemming out of the self-consistent
action-angle transport theory is  \cite{GattoOPPJ2011}

\bea
{\cal V}_a
\frac{\partial \left\langle M N({\bf x};t) V_\para({\bf x};t)
\right\rangle_{\bar{r}}}{\partial t}
&-& {\cal V}_a \langle q_1 N  E_t \rangle_{\bar{r}}
+ \frac{\partial }{\partial \bar{r}} \langle {\cal V}_a V_f M N V_\para
\rangle_{\bar{r}} ~S(1 - \kappa)
\nonumber \\
&+&  \frac{\partial}{\partial \bar{r}}
{\cal V}_a  \Gamma^V(\bar{r})
= (\bar{r} B_{0,t}) {\cal V}_a  U^V~,
\label{fl:5.2.4-1}
\eea
where
${\cal V}_a = 4 \pi^2 R_0 w \bar{r}$ is the toroidal shell  of width $w$ (centered at
$\bar{r}$) over which magnetic perturbation mode $a$ generated by the bulk ions is nonzero, $E_t = (1/c R) (\partial \psi_p
/ \partial t)$ is the induced toroidal electric field
at a fixed position in
space, while $\Gamma^V$ and $U^V$ account for the momentum transport and generation due to
fluctuations, respectively.
The third term on the LHS of Eq.~(\ref{fl:5.2.4-1}) containing the velocity of a flux
surface, $ V_f=c E_t/B_p$, is a collisionless version of the
Ware-Galeev pinch, effective only for trapped
particles as described by the step function $S(1 - \kappa)$
yielding $1$ for trapped particles ($\kappa\in [0,1]$) and $0$ for circulating
($\kappa\in (1,+\infty)$). Eq.~(\ref{fl:5.2.4-1}) is a generalization of the quasilinear result of
Ref.~\cite{Mahajan83}.

The momentum flux and the momentum source present in Eq.~(\ref{fl:5.2.4-1})
have been derived in
Ref.~\cite{Gatto07,Gatto06} and Ref.~\cite{Chavdarovski09}, respectively, with
\bea
\Gamma_{ei}^V
&=&- {\cal L}_{ei} \hat{\Gamma}^{ei}_2  M_e \frac{d V_{\para e}}{d r}
- {\cal L}_{ei} \hat{\Gamma}^{ei}_1
 \frac{M_e V_{\para e}}{\rho_{\rm e,p}}
 \label{Gamma_eiV}
\eea
and
\bea
(r B_{0t}) U_{ei}^V
&=&
- {\cal L}_{ei} \hat{U}^{ei}_2
\frac{M_e}{\rho_{\rm e,p}} \frac{d V_{\para e}}{d r}
- {\cal L}_{ei} \hat{U}^{ei}_1 \frac{M_e V_{\para e}}{\rho_{\rm e,p}^2}
~,\label{U_eiV}
\eea
where the following non-dimensional radial functions were introduced
\bea
&&
\!\!\!\!\!\!\!\!\!\!\!\!\!\!\!\!\!\!\!\!\!\!\!\!\!\!\!\!\!\!\!\!\!
\hat{\Gamma}^{ei}_1
=
3 \rho_{\rm e,p}
\left[
\frac{1}{p_e} \frac{d p_e}{dr} + \frac{1}{p_i} \frac{d p_i}{dr}
- 2 \left( \frac{T_i}{T_e} - \frac{3}{2} \right)
 \left( \frac{1}{q_{\rm saf}} \frac{d q_{\rm saf}}{d r} \right)
\right]~,~~~~~
\hat{\Gamma}^{ei}_2
= 3, \label{Gamma_1_2} \\
&&
\!\!\!\!\!\!\!\!\!\!\!\!\!\!\!\!\!\!\!\!\!\!\!\!\!\!\!\!\!\!\!\!\!
\hat{U}^{ei}_1
=
 15 \rho_{\rm e,p}^2
\left( \frac{1}{q_{\rm saf}} \frac{d q_{\rm saf}}{d r} \right)^2
\! \frac{T_i}{T_e} \!
+ 3 \rho_{\rm e,p}^2
\left( \frac{1}{q_{\rm saf}} \frac{d q_{\rm saf}}{d r} \right)
\frac{T_i}{T_e}
\left(
\frac{1}{N_e} \frac{d N_e}{dr} + \frac{1}{N_i} \frac{d N_i}{dr}
+ \frac{2}{T_i} \frac{d T_i}{dr}
\right) , \label{U_1} \\
&&
\!\!\!\!\!\!\!\!\!\!\!\!\!\!\!\!\!\!\!\!\!\!\!\!\!\!\!\!\!\!\!\!\!
\hat{U}^{ei}_2
=
 3 \rho_{\rm e,p}
\left( \frac{1}{q_{\rm saf}} \frac{d q_{\rm saf}}{d r} \right)
\frac{T_i}{T_e}~.
\label{U_2}
\eea
Here the electron poloidal gyro-radius and gyro-frequency
are defined by, respectively,
$\rho_{\rm e,p} = v_{\rm th,e}/\Omega_{\rm e,p}$ and
$\Omega_{\rm e,p} = e B_{0,\theta}/(c M_e)$.
The transport coefficient is given by
${\cal L}_{12} = \sum_{r_a} p^2 \pi N_1 b_t^2
D_{RR}(1,2)$,
where
$D_{\rm RR}(1,2)$
is a generalized Rechester-Rosenbluth
coefficient \cite{Rechester78}.
In Eq.~(\ref{fl:5.2.4-1}) we only keep the electron flux and source
due to the fluctuation spectrum induced by the ions,
as shown by the subscript ``ei''. The
electron-electron interaction $\Gamma_{ee}^V$ and $U_{ee}^V$
is neglected due to the use of the pseudo-thermal ansatz \cite{Mynick89} that makes
these terms equal to zero when volume-integrated. However, these terms are
locally nonzero and
in order to have a better estimate of their effect the use of a
less restrictive turbulent generalized Balescu-Lenard (gBL)
collision operator would be required, one obtained through a
self-consistent phenomenological
evaluation of the turbulent spectrum.

The first terms in Eqs.~(\ref{Gamma_eiV}) and~(\ref{U_eiV}) are diagonal terms
representing MHD effects, while the others are off-diagonal, purely kinetic terms.
The existence of off-diagonal terms in the momentum
transport matrix has been predicted theoretically and confirmed experimentally~\cite{Ida95}.
The drive in the off-diagonal terms is related to ion and electron density and
temperature gradients,
as well as the gradient of the safety factor $q_{\rm saf}$.
The source terms in Eqs.~(\ref{U_1}) and~(\ref{U_2})
are proportional to the safety factor gradient, meaning the
momentum source will only play a role when the magnetic shear is large.
Differently, in Eq.~(\ref{Gamma_1_2}) the safety factor term
contains the $T_i/T_e$ ratio, and can change the direction of the momentum flux
depending on the electron and ion temperatures and on the magnetic configuration.
In conventional scenarios with $q_{\rm saf}$ monotonically
increasing with $r$, a momentum pinch is obtained for
$T_i/T_e < 3/2$.

The turbulent electric field
is derived from the electron momentum balance Eq. (\ref{fl:5.2.4-1}),
where the flux and the source are given in
Eqs.~(\ref{Gamma_1_2})-(\ref{U_2}).
Assuming for simplicity the ion drift velocity is negligible,
$V_{\para i} = 0$, and approximating the electron
velocity as $V_{\para e} \simeq - j_\para /
(e N_e)$, Eq.~(\ref{fl:5.2.4-1}) for the passing-electron gives
\beq
- \frac{M_e}{e^2 N_e} \frac{\partial \left\langle j_{\para}
\right\rangle_{r}}{\partial t}
+ \frac{\left\langle N_e E_{t} \right\rangle_{r}}{N_e}
= \eta_{\rm neo} j_\para - E_{\rm bs} + E_\para^{\rm turb}~.
\label{Ohm_law_turb}
\eeq
The RHS of Eq.~(\ref{Ohm_law_turb}) contains two additional terms, the
neoclassical resistive term $\eta_{\rm neo} j_\para$
and the neoclassical bootstrap term, $E_{\rm bs} \propto - d p / d r$, where $p$ is the total equilibrium
pressure. The turbulent electric field is then written in the following form
\bea
E_\para^{\rm turb}
= \eta_{\rm an} j_\para + \eta_\times \frac{d j_\para}{d r}
- \frac{1}{B_0 r} \frac{d}{dr} \left[ r \eta_{\rm h} \frac{d}{dr}
\left( \frac{j_\para}{B_0} \right) \right]~
\label{E_SC}
\eea
in order to make connection to the MHD framework
of mean-field \cite{Boozer86,Bhattacharjee86}, in which case
hyper-resistivity $\eta_{\rm h}$ is equivalent to the $\alpha$-term in the dynamo theory, while
anomalous resistivity $\eta_{\rm an}$ to the $\beta$-term~\cite{Moffat78}. In Eq.~(\ref{E_SC}) $B_0$ is the total equilibrium magnetic field,
and the three transport coefficients (anomalous resistivity, cross-
resistivity and hyper-resistivity) are given by
\beq
\eta_{\rm an}
\equiv
- \frac{1}{B_0 r} \frac{d}{dr} \left( \frac{r \Lambda_2^\Gamma}{N_e}
\frac{d B_0}{dr} \right) + \frac{\Lambda_1^\Gamma + r ( d \Lambda_1^\Gamma
/ dr)}{N_e r} - \Lambda_1^U~,
\label{an_res}
\eeq
\beq
\eta_\times
\equiv - \frac{\Lambda_2^\Gamma}{N_e L_{N_e}}
+ \frac{\Lambda_1^\Gamma}{N_e}
+ \Lambda_2^U ~,
\label{sta_cross}
\eeq
and
\beq
\eta_{\rm h} \equiv \Lambda_2^\Gamma B_0^2 / N_e~.
\label{hyp}
\eeq
To shorten the previous equations we have used the notations:
\beq
\Lambda_1^\Gamma = \frac{\eta_{ei}}{\rho_{\rm e,p}} \left(
\hat{\Gamma}^{ei}_2 \frac{\rho_{\rm e,p}}{L_{N_e}} - \hat{\Gamma}^{ei}_1
\right)~,~~~\Lambda_2^\Gamma = \eta_{ei} \hat{\Gamma}^{ei}_2~,
\label{lambda_a}
\eeq
\beq
\Lambda_1^{\rm U} =  \frac{1}{N_e \rho_{\rm e,p}}
\frac{\eta_{ei}}{\rho_{\rm e,p}} \left(
\hat{U}^{ei}_2 \frac{\rho_{\rm e,p}}{L_{N_e}} - \hat{U}^{ei}_1
\right)~,~~~\Lambda_2^U = \frac{1}{N_e \rho_{\rm e,p}}
\eta_{ei} \hat{U}^{ei}_2 ~,
\label{lambda_b}
\eeq
where $L_{N_e}^{-1} \equiv (1/N_e) (d N_e/dr)$, and the transport coefficients are
$\eta_{ei} = M_e {\cal L}_{ei} / (e^2 N_e)$ and
${\cal L}_{ei}/N_e =  \pi v_{th,e} q_s R_0 \tilde{b}^2$, with $\tilde{b}$ the normalized
(to $B_0$) magnetic perturbation.

\section{Turbulent transport coefficients}
\label{coeff}

The last term in Eq.~(\ref{E_SC}), hyper-resistivity $\eta_{\rm h}$, originates
from the momentum flux. It is proportional to the anomalous viscosity
and has a tendency to smooth out the radial gradient of the parallel
current density near the axis, which could explain why experimental current profiles
remain non-hollow even in the presence of strong bootstrap current \cite{Hwang96}.
The magnetic turbulence profiles in this paper are chosen to vanish at the boundary, making
the hyper-resistive term helicity conserving.
Eq.~(\ref{E_SC}) shows that in a Taylor state \cite{Taylor74}, $j_\para / B_0 = {\rm constant}$,
the effects of this term are vanishing, so we can say that the hyper-resistive term is
driven by departures from a Taylor state. This term is also responsible for
toroidal field reversal at the plasma edge in pinches \cite{Bhattacharjee86}.
The remaining two coefficients, the anomalous resistivity Eq.~(\ref{an_res}) and cross-resistivity Eq.~(\ref{sta_cross})
contain both momentum flux and momentum source contributions, a result of the self-consistency of our theory.
Hyper-resistivity is always a dissipative term, while contrary to that the cross-resistivity
usually generates power from the turbulence. The anomalous term is mostly dissipative, but we show that this can change
depending on the temperature and perturbation profile.

\begin{figure}
\epsfxsize=0.7\linewidth
\centerline{\epsfbox{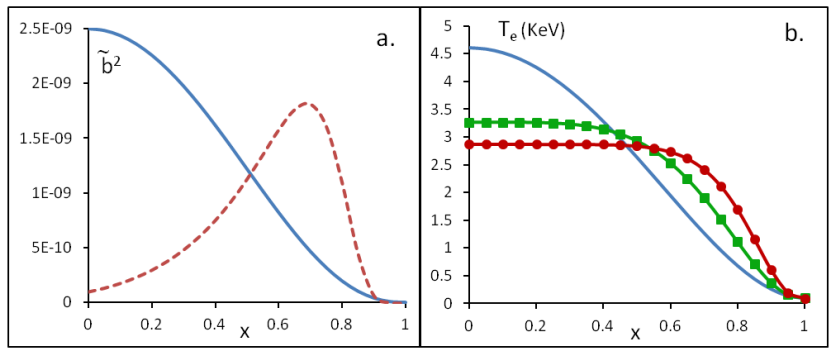}}
\caption{{\rm a}. Profiles of perturbation $\tilde{b}^2$: full line-peaked at the axis,
dashed line- peaked off-axis; {\rm b}. Electron temperature profiles: full line- peaked L-mode,
boxes- broad L-mode, circles- advanced scenario. \label{fig:profile}}
\end{figure}
In this section we present the radial dependence of
the three turbulent transport coefficients
Eqs.~(\ref{an_res})-(\ref{hyp}) as functions of the temperature and turbulence
perturbation profile.
For the tokamak dimensions we take $a =71$ cm, $R_0 =240$ cm and obtain an inverse aspect ratio of
$\epsilon \equiv a/R_0 = 0.295$. The electron (and ion) density profile is given by
$
N_e=N_i=(N_{e,0}-N_{e,a}) (1-x^2)^{\gamma_N}+N_{e,a}\, ,
$
where $x = r/a$, $\gamma_N=1.5$, $N_{e,0}=2 \times 10^{13}$ 1/cm$^3$, and
$N_{e,a}= 2\times 10^{11}$  1/cm$^3$ for all simulations in this paper.
The electron and ion temperature profiles are given with the function
$
T_e=T_i=(T_{e,0}-T_{e,a}) (1-x^{\sigma_T})^{\gamma_T} + T_{e,a},
$
where $T_{e,0}$, $T_{e,a}$, $\sigma_T$ and $\gamma_T$ are parameters to be chosen.
We will compare the turbulent coefficients for
two L-modes shown in Fig.\ref{fig:profile}b, one peaked L-mode (full line) and one broad (boxes).
Additionally, for  comparison in the same figure we show the advanced-like mode (circles)
that will be studied in Sec. \ref{advanced}.
For the peaked L-mode we take
$\gamma_T=2.0$, $\sigma_T=2.0$ with boundary values
$T_{e,0}= 4.606$ keV and
$T_{e,a}= 0.1 $ KeV.
The broad mode is obtained for
$\gamma_T=2.5$ and $\sigma_T=4.5$, with boundary values
$T_{e,0}= 2.875$ keV and $T_{e,a}= 0.09375 $ keV.
This mode has a strong gradient in the outer region,
and hence the bootstrap current is expected to be
wider compared to the one of the peaked mode.
The advanced mode is given by the parameters
and  $\gamma_T=3.0$, $\sigma_T=8.0$,
$T_{e,0}= 2.87$ keV and $T_{e,a}= 0.0878$ keV.
This mode has an even steeper pressure gradient near the edge of the plasma.
For all profiles
in this section we will take $T_e=T_i$, and hence $p_e=p_i=p/2$.

In Ref.~\cite{GattoOPPJ2011}
simulation was done with magnetic perturbation $\tilde{b} \sim O(10^{-4})$ peaked at the center.
This level of turbulence is rarely present at the axis where turbulence is
vanishing due to flat profiles, but it's not uncommon at the edge.
While density
perturbations are experimentally well measured, the same
cannot be said about magnetic perturbations due to the small amplitude and lack of good diagnostics.
Experiments with heavy ion probing at the center in JIPPT-
IIU tokamak show turbulence level of $O(10^{-4})$~\cite{Hamada15}, which however is not expected to be
present at the axis of larger machines. Measurements off-axis in Tore Supra using cross-polarization
scattering of microwaves~\cite{Colas98} report similar levels in L-mode.
The level of turbulence we examine in this paper ($\tilde{b} \sim O(10^{-5}-10^{-4})$) provides for the
ergodization of
the magnetic surface which causes for the electrons streaming along the
magnetic field lines to execute a random-walk in a stochastic magnetic field~\cite{Rechester78}.
Contrary to that the MHD approach to this problem requires small amplitude MHD fluctuations,
and not necessarily stochastic field. This is not
a limitation of the kinetic theory since some stochastization of the field is present even
for significantly lower levels of microturbulence $\tilde{b}\sim O(10^{-7}-10^{-6})$~\cite{Firpo15}.

To understand better the effects of the turbulence
we assume the presence of densely-packed micro-tearing
modes described by two different (intermediate level)
perturbation profiles: one peaked at the axis (full line in Fig.\ref{fig:profile}a) given by
$\tilde{b}^2=0.25 \times 10^{-8} (1-x^2)^{2.5}$, and one for
turbulence near the edge
with $\tilde{b}^2=10^{-10} \times (1-x^{10})^{10}(1+x)^{6}$
(dashed line in Fig.\ref{fig:profile}a). For completeness we will also show results for
stronger turbulence $\tilde{b}^2=O(10^{-8})$ at the center, and weaker $\tilde{b}^2 \lesssim O(10^{-10})$
everywhere in the plasma. For a more accurate analysis, in future works we will compare our results
with turbulence profiles obtained from more consistent numerical codes.
All perturbations in this work are taken to be zero at the plasma
edge, leading to a helicity-conserving hyper-resistivity. Since, at
the moment we are more interested in the core plasma
this assumption is justified, even though it neglects the proper treatment of edge current diffusion and
possibly the effects of turbulence on
impurity transport.
All turbulent coefficients further are normalized with
$\bar{\eta}_{\rm cl}$, the cylindrical cross-section average of the
classical resistivity
$
\eta_{\rm cl} = 4.77 e^2 Z_{\rm eff}
\ln \Lambda/[M_e v_{{\rm th},e}^3]
$,
where the Coulomb logarithm and the ion charge are
$\ln \Lambda = 17$ and $Z_{\rm eff}=1$, respectively.
To avoid the singularity of the
neoclassical resistivity
$
\eta_{\rm neo} = \eta_{\rm cl}/(1 - 1.95 \sqrt{r/R_0})
$ we will simply take $\eta_{\rm neo} = \eta_{\rm cl}$,
since this issue is not essential to our study.
\begin{figure}
\epsfxsize=0.7\linewidth
\centerline{\epsfbox{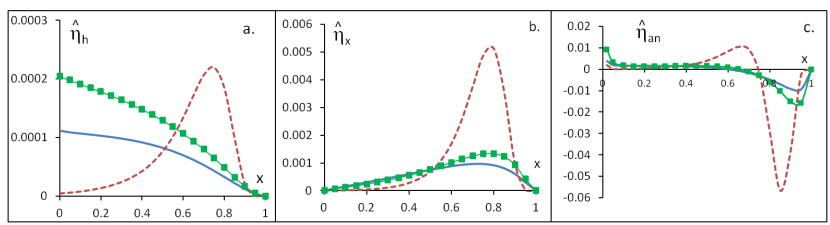}}
\caption{Normalized turbulent coefficients; {\rm a}. Hyper-resistivity, {\rm b}. Cross-resistivity,
{\rm c}. Anomalous resistivity;
Full line is for peaked mode and axis turbulence, dashed line for peaked mode and edge turbulence,
boxes for broad mode and axis turbulence.
\label{fig:koef}}
\end{figure}
The non-dimensional hyper-resistive coefficient
$\hat{\eta}_{\rm h} = \eta_{\rm h} / (a^2 B_{z,0}^2 \bar{\eta}_{\rm cl})$,
cross coefficient
$\hat{\eta}_{\rm X} = \eta_{\rm X} / (a \bar{\eta}_{\rm cl})$, and the
anomalous coefficient
$\hat{\eta}_{\rm an} = \eta_{\rm an}/\bar{\eta}_{\rm cl}$
are plotted in Fig.~\ref{fig:koef} (a)-(c), respectively, for two different temperature and
turbulent profiles. In Fig.~\ref{fig:koef} the full lines represent a peaked mode with axis turbulence,
dashed lines- peaked mode with edge turbulence, and boxes- broad mode with axis turbulence.
All coefficients go to zero at the edge of the plasma, due to the adopted shape of the magnetic
perturbation, while in the inner region they are all sensitive to the equilibrium ($N, T$) and
turbulence ($\tilde{b}^2$) profiles. Hyper-resistivity is proportional
to $\tilde{b}^2$ (see Eq.~(\ref{hyp})) and takes on the shape of the perturbation, while the
shapes of the other two do not change in a
significant way when switching from axis to edge turbulence. However,
the values of all three parameters in this situation decrease near the axis,
but significantly increase in the outer region. This is
due to the alignment of the turbulence peak with the temperature gradient.
The coefficients are also generally larger for a broader mode.

While the hyper- and the cross-resistive contribution are
positive everywhere in the plasma, the anomalous resistivity switches sign from positive
to negative when going to large $x$, depending on equilibrium thermodynamic
and magnetic ($q_{\rm saf}$) profiles, as Eqs.~(\ref{Gamma_1_2})-(\ref{U_2}) show.
This term is on average larger than the other two, but it's
still significantly smaller than the classical resistivity.
In Eq.~(\ref{an_res}) we have written $\hat{\eta}_{\rm an}$ as a sum of three terms, out of which the
second one is shown to be dominant~\cite{GattoOPPJ2011}.
This coefficient has a removable singularity at the axis (see Eq.~(\ref{an_res})) that disappears when the Ohm's law is
iteratively solved due to the readjustment of the magnetic profile. The existence of negative anomalous
resistivity region has already been postulated in several works using the MHD approach \cite{Biskamp84}.
In section \ref{advanced} we show that in advanced scenario anomalous
resistivity can generate power instead of dissipating it, and this to the best of our knowledge is a novel result.

\section{Ohm's law and power balance}
\label{balance}

The steady state version of Eq.~(\ref{Ohm_law_turb}) can be written as
$E_\para^0 = \eta_{\rm neo} \left( j_\para - j_{\rm bs} \right) +
E_\para^{turb}$
or, using Eq.~(\ref{E_SC}),
\beq
E_\para^0 =
-  \eta_{\rm neo} j_{\rm bs}
+ (\eta_{\rm neo} + \eta_{\rm an} ) j_\para
+ \eta_\times \frac{d j_\para}{d r}
- \frac{1}{B_0 r} \frac{d}{dr} \left[ r \eta_h \frac{d}{dr}
  \left( \frac{j_\para}{B_0} \right) \right]~,
\label{ohmsteady1}
\eeq
where for the
neoclassical bootstrap current we adopt \cite{Rawls75,Hinton72,Hazeltine73,Pfeiffer80,Yuan93}:
\bea
&&
\!\!\!\!\!\!\!\!\!\!\!\!\!\!\!\!\!\!\!\!\!
j_{\rm bs}
=
-  c F_{13} \left( \frac{r}{R} \right)^{1/2} \frac{n_e(T_e+T_i)}
{B_\theta} \left( \frac{1}{n_e} \frac{d n_e}{dr} \right)
- c \left( \frac{r}{R} \right)^{1/2} \frac{n_e T_e}{B_\theta}
\nonumber \\
& &
\!\!\!\!\!\!\!\!\!\!\!\!
\times
\left[ - \left( \frac{3}{2} F_{13} - F_{23} \right) \left( \frac{1}{T_e}
\frac{d T_e}{dr} \right)
- \left( \frac{3}{2} - y \right) F_{13}
\frac{T_i}{T_e}
\left( \frac{1}{T_i} \frac{d T_i}{dr} \right) \right]~. \label{Jneo}
\eea
Here
\[
F_{m3} = \frac{K_{m3}}{[1+a_{m3} \nu_{e *}^{1/2}+b_{m3} \nu_{e *}]
[1+c_{m3} \nu_{e *} (r/R)^{3/2}]
}~,
~~~~m=1,2~,
\]
\[
y = \frac{1.31 (1 + 1.65 \nu_{i *}^{1/2})}{1+0.862 \nu_{i *}^{1/2}}~,
\]
\[
\nu_{j *} =  \frac{B_z R^{3/2}}{\tau_j r^{1/2} B_\theta (T_j/m_j)^{1/2}}~,
~~~~~j=e,i~,
\]
the classical e-i and i-i collision times are
$\tau_e = 3 m_e^{1/2} T_e^{3/2} / [ 4 (2 \pi)^{1/2} \ln \Lambda_e n_e e^4
Z_{\rm eff}]$,
$\tau_i = 3 m_i^{1/2} T_i^{3/2} / [ 4 \pi^{1/2} \ln \Lambda_i n_i e^4 ]$,
and for $Z_{\rm eff}=1$ we have $K_{13}=2.30$, $K_{23}=4.19$, $a_{13}=1.02$, $a_{23}=0.57$,
$b_{13}=0.75$, $b_{23}=0.38$, $c_{13}=1.07$ and $c_{23}=0.61$
\cite{Pfeiffer80}.
The frequency $\nu_*$ is a measure of the collisionality of the plasma and is given by
the ratio of the effective collision frequency to the bounce frequency.
There are three main collisionality regimes: $ \nu_* \ll 1$ banana (collisionless) regime,
$1 \leq \nu_* \leq 1 / \epsilon^{3/2}$ plateau and $\nu_* \gg 1$ Pfrisch-Schl$\ddot{\rm u}$ter
(collisional) regime. Here, we obtain an average value of $\nu_{e *} \approx 0.05$,
which falls in the banana regime, as is necessary for maintaining a bootstrap current.
Since the poloidal field $B_{\theta}$ appears in the denominator
of the expression for the bootstrap
current Eq.~(\ref{Jneo}) and is connected to the parallel
current density via Ampere's law, $(4 \pi/c) j_\para \simeq (1/r) (\partial/\partial r)
r B_{\theta}$, Eq.~(\ref{ohmsteady1}) is a nonlinear integro-differential equation and is
therefore solved iteratively.

Ohm's law Eq.~(\ref{ohmsteady1})
represents a steady-state balance between charged particles
momentum gain and momentum loss expressed through the parallel
current density profile~\cite{GattoOPPJ2011}.
Solutions of this equation can explain whether or not the
turbulent contributions of Ohm's law can provide the
necessary current diffusion towards the axis in operations
with large bootstrap component \cite{Weening92,Itoh93,Yuan93}.
Our work shows that all three terms play significant role in shaping the
current and safety factor profile in presence of magnetic turbulence,
and hence must be retained in the analysis.
A complete study of the effects of these terms
on the evolution of plasma equilibrium should include a
transport code which couples the turbulent Ohm's law
to the time dependent equations for the temperature and density.
Additionally, the turbulence profile and intensity would also change self-consistently.
Here, we will concentrate on solving Eq.~(\ref{ohmsteady1})
for a steady-state current profile in a (cylindrical) tokamak
with fixed pressure equilibrium profiles, and for a fixed (in time) level of magnetic
turbulence.

We can further investigate each resistive term using the power balance equation
\bea
& &
\!\!\!\!\!\!\!\!\!\!\!\!\!\!\!\!\!\!\!\!\!\!\!\!\!\!\!\!\!\!\!\!\!\!\!\!\!\!
\!\!\!\!\!
\underbrace{\int_0^a \!\!dr~r~j_\para E_\para^0}_{{\cal T}_{\rm E}}
+ a \left\{ \eta_h  \frac{j_\para }{B_0}
\frac{d}{dr}  \left[  \frac{j_\para}{B_0} \right]
- \frac{1}{2} \eta_\times j_\para^2 \right\}_{r=a}
\nonumber \\
&=& \underbrace{\int_0^a \!\! dr~[\eta_{\rm neo} + \eta_{\rm an} ] r~j^2_\para}_{{\cal T}_{\rm neo}+{\cal T}_{\rm an}}
+
 \underbrace{\int_0^a \!\! dr~r~\eta_h \left[
\frac{d}{dr}  \left(  \frac{j_\para}{B_0(r)} \right) \right]^2}_{{\cal T}_{\rm h}}
\nonumber \\
&+& \underbrace{\int_0^a \!\!dr~r~j_\para \left[ -\eta_{\rm neo} j_{\rm bs}
- \frac{1}{2 r} \frac{d [r \eta_\times]}{d r} j_\para \right]}_{{\cal T}_{\rm BS}+{\cal T}_{\rm \times}}
~, \label{power_balance}
\eea
obtained by multiplying Eq.~(\ref{ohmsteady1}) with $rj_\para$ and
then integrating from $0$ to $a$, according to the procedure adopted in
Ref.~\cite{Itoh93,GattoOPPJ2011}.
The first term on the LHS is positive and describes the externally injected power by the
induced electric field $ E_\para^0$.
The second term is the power injected from the boundary surface
($r=a$) of the plasma which in this case is equal to $0$ due to the mentioned conservation
of helicity ($j_\para(a) \simeq 0$).
The first term on the RHS shows the
internal dissipation due to neoclassical resistivity
and dissipation (or generation) due to anomalous effects. Fig.~\ref{fig:koef}c shows that
the anomalous resistivity is negative (and generates current) on the outer region of the plasma.
However, the current density is generally small on the same region, so that in most common cases the overall effect
of the anomalous term is dissipation. This however can change when there is an alignment of the region
of negative
anomalous resistivity with strong plasma pressure gradient (see Sec.\ref{advanced}).
Most of the power generated internally in all operations comes
from the diffusion-driven (bootstrap) electromotive force, reinforced
by the turbulent contribution from $\eta_{\rm \times}$.
It is shown in section \ref{fully} that the total bootstrap term ${\cal T}_{\rm BS}+{\cal T}_{\rm \times}$
could balance the dissipative terms on the RHS, eliminating the need of externally
supplied power.
Finally, the second term on the RHS, represents
the additional power (always) dissipated by the hyper-resistive current
diffusion.

\section{Numerical solution of the current equation}
\label{numerical}
When numerically solving Eq.~(\ref{ohmsteady1})
we will use $x \equiv r/a$ as a radial variable and put the relevant
quantities in dimensionless form (marked with an over-hat):
$\hat{j}_\para = 4 \pi a j_\para / (c B_{z,0})$,
$ \hat{E}_\para^0 = 4  \pi a E_\para^0/ (\bar{\eta}_{\rm cl} c B_{z,0})$,
$\hat{P} = P / B_{z,0}^2$,
$\hat{B} = B/B_{z,0}$,
$\hat{\eta}_{\rm neo} = \eta_{\rm neo} / \bar{\eta}_{\rm cl}$,
$\hat{\eta}_{\rm h} = \eta_{\rm h} / (a^2 B_{z,0}^2 \bar{\eta}_{\rm cl})$,
$\hat{\eta}_\times = \eta_\times / (a \bar{\eta}_{\rm cl})$,
$\hat{\eta}_{\rm an} = \eta_{\rm an}/\bar{\eta}_{\rm cl}$,
where $\bar{\eta}_{\rm cl}$ is the volume averaged classical resistivity
and $B_{z,0}=4 \times 10^4$ G
is the toroidal field.
The final form of the parallel current equation is then
\beq
\!\!\!\!\!\!\!\!\!\!\!\!\!\!\!\!\!\!\!\!\!\!\!\!\!
\hat{D}_2(x)
\frac{d^2 \hat{j}_\para(x)}{d x^2}
+
\hat{D}_1(x)
\frac{d \hat{j}_\para(x)}{d x}
+
\hat{D}_0(x)
\hat{j}_\para(x)
= \frac{\hat{D}_{-1}(x)}{\int^x d x^\p~x^\p~\hat{j}_\para(x^\p)}
+ \hat{E}_\para^0(x)~,
\label{final_eq_jpara}
\eeq
with
\bea
&&
\!\!\!\!\!\!\!\!\!\!\!\!\!\!\!\!\!\!\!\!\!\!\!\!\!
\hat{D}_2(x;\eta_{\rm h})
= - \frac{\hat{\eta}_{\rm h}}{\hat{B}_0^2} \label{D2}
~,~ \hat{D}_1(x;\eta_{\rm h},\eta_\times)
=
\hat{\eta}_\times - \frac{1}{\hat{B}_{0,z}^2}
 \frac{1}{x} \frac{d ( x \hat{\eta}_{\rm h} )}{dx}
~, \nonumber \\
&&
\!\!\!\!\!\!\!\!\!\!\!\!\!\!\!\!\!\!\!\!\!\!\!\!\!
\hat{D}_0(x;\eta_{\rm neo},\eta_{\rm an})
= \hat{\eta}_{\rm neo} + \hat{\eta}_{\rm an}
~,\nonumber \\
&&
\!\!\!\!\!\!\!\!\!\!\!\!\!\!\!\!\!\!\!\!\!\!\!\!\!
\hat{D}_{-1}(x;\eta_{\rm neo})
=
- 4 \pi \hat{\eta}_{\rm neo} \left( \frac{a}{R_0} \right)^{1/2}
x^{3/2} \nonumber \\
&&
\times \left\{ F_{13} \frac{n_e [T_e+T_i]}{B_{z,0}^2}
\left( \frac{1}{n_e} \frac{d n_e}{dx} \right)
+\frac{n_e T_e}{B_{z,0}^2}
\right. \nonumber \\
&&
\left. \times
\left[ - \left( \frac{3}{2} F_{13} - F_{23} \right)
\left( \frac{1}{T_e} \frac{d T_e}{dx} \right)
- \left( \frac{3}{2} - y \right) F_{13} \frac{T_i}{T_e}
\left( \frac{1}{T_i} \frac{d T_i}{dx} \right) \right] \right\}~,
\label{Dm1}
\eea
where $\hat{D}_1$ and $\hat{D}_0$ were expanded to the lowest order term of the
small parameter $\epsilon \sim B_{0,p}/B_{0,z} \sim a / R_0$\footnote{We would like to correct a typographical error in Ref.~\cite{GattoOPPJ2011} where in Eq.~($39$) the derivatives should be with respect to $x$, not $r$.}.

A numerical code, Turbulent Ohm's Law Solver (abbr. TOLS) iteratively solves the
nonlinear Eq.~(\ref{final_eq_jpara})
by taking the Ohmic+Bootstrap current as initial solution and recalculating the fields, safety factor,
coefficients and the current when turbulence is added, until a converged solution is achieved.
The effect of the turbulent terms limited to a peaked L-mode profile have been studied in Ref.~\cite{GattoOPPJ2011}.
To better understand the importance of each term, in this paper we use several different scenarios and two
different turbulence profiles. For simplicity sake we take the ion and electron temperatures to be equal in all scenarios,
except for the fully bootstrapped tokamak in Sec.~\ref{fully}, which includes an electron transport barrier.
As previously stated we
approximate $\hat{\eta}_{\rm neo} \simeq \hat{\eta}_{\rm cl}$, and assume the boundary
condition $\hat{j}_\para(x=1)=0$ in all cases, while additional condition $\hat{j}_\para^\p(x=0)=0$ is applied when the hyper-resistivity is taken into account. The normalized inductive field is $\hat{E}_\para^0=7.0625 \times 10^{-3}$.
All parameters are chosen so that the stored energy is equal in all scenarios, while the current inside the plasma (Ohmic + Bootstrap) is $700$KA.

\subsection{L-mode type profiles}
\label{L-mode}

\begin{figure}
\epsfxsize=0.7\linewidth
\centerline{\epsfbox{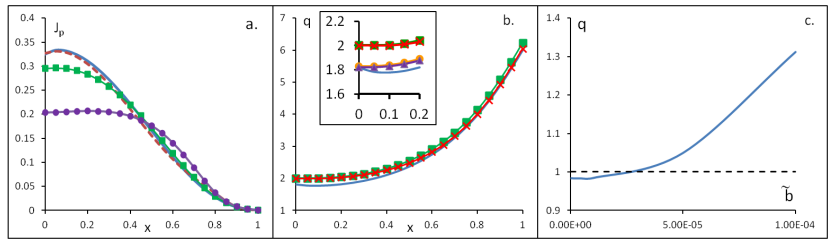}}
\caption{{\rm a}. Parallel current of the peaked L-mode. Full line is for Ohmic+Bootstrap
current only, boxes- with axis turbulence, dashed line- with edge turbulence, circle- $O(10^{-4})$ axis turbulence. {\rm b}.
Safety factor of the peaked L-mode with axis turbulence. Full line- Ohmic+Bootstrap
current only, circles- added hyper-resistivity, triangles- hyper + cross-resistivity, boxes-
 hyper + anomalous resistivity, crosses- all coefficients. {\rm c}. Safety factor at the axis vs. $\tilde{b}$.
\label{fig:Lmode1}}
\end{figure}

In this section we will consider the two temperature profiles from Fig.~\ref{fig:profile}b, the peaked mode given with full line
and the broad mode given with boxes. In the peaked L-mode temperature  profile, the bootstrap current
constitutes $14.6\%$ of the initial current. In Ref.~\cite{GattoOPPJ2011} it was showed that the hyper-resistivity, although the smallest among the coefficients, plays an important role near the axis by diffusing the current density toward the center and
hence flattening the current profile. This effect is essential in dealing with the
hollowness induced by the bootstrap current in advanced scenarios (see sections ~\ref{advanced} and ~\ref{fully}).
Checking first the effect of each single turbulent term in isolation, using the on-axis turbulence profile of Fig.~\ref{fig:profile}a, we find that the hyper-resistivity reduces the current by $1.2\%$. The anomalous resistivity reduces the current in the central region of the plasma,
but also increases it in the outer region due to the negative value of $\hat{\eta}_{\rm an}$,
with the overall effect being reduction by roughly $2.5\%$. The cross-resistivity, on the other hand increases the current everywhere in the plasma with total amplification by some $3\%$.
We explicitly note that the increases/reductions of the current due to the turbulent terms do not simply add, but combine according to a complicated differential equation.

When all turbulent coefficients are included we obtain the current profiles in Fig~\ref{fig:Lmode1}a, where the full line is for Ohmic+Bootstrap current only, boxes- with axis turbulence, dashed line- with edge turbulence, circles- for strong turbulence  $O(10^{-4})$ on axis. For moderate level of turbulence, the final effect of all turbulent terms is reduction of the current by $1\%$, which means the total current before and after inclusion of the turbulence is for all practical purposes unaltered.
However, we see from Fig.~\ref{fig:Lmode1}a (boxes) that the turbulence internally redistributes
about $5.5\%$ of the total current, or in absolute values, $2.43 A/cm^2$. In this scenario two terms generate power internally
(given in relative values),
${\cal T}_{\rm \times} = 3.2\%$ and ${\cal T}_{\rm BS} = 15.8\%$, with the rest generated
inductively, while the other terms dissipate the power as follows ${\cal T}_{\rm h} = 1.1\%$,
${\cal T}_{\rm an} = 2.6\%$ and the majority by the neoclassical term.
When the turbulence is peaked off-axis, the current change around the axis is negligible, but there is some
current change in the outer region (Fig.~\ref{fig:Lmode1}a - dashed line), and due to the competing effects of the anomalous and cross-resistivity, a small current reduction of $1\%$. When turbulence is strong at the center ($O(10^{-4})$- shown with circles) there is still a small reduction of $4\%$, but a significant current redistribution, while for $\tilde{b}<10^{-5}$ everywhere in the plasma the effects become negligible (order $0.1\%$ of the total current). For intermediate level of turbulence the redistribution of the current is enough to cause the $q_{saf}$ profile to raise (see Fig.~\ref{fig:Lmode1}b). This is important in discharges with $q_{saf}<1$ on axis, where raising the value to $q_{saf}>1$ will prevent a sawtooth crash.
A paper on one such case from ASDEX Upgrade is in preparation.

Several works have considered the existence of sawtooth-free hybrid discharges
and credited the stability of the stationary state to
several different phenomena: hyper-resistivity~\cite{Casper07}, rotating island driving drift current~\cite{Chu07},
critical poloidal current density~\cite{Garcia10}, $3/2$ tearing mode~\cite{Petty09},
fishbone activities~\cite{Wolf99,Gunter99}, the exact effect of which on the safety factor is yet to be clarified.
In a simulation using M3D-C1 (3D resistive MHD) code~\cite{Jardin15}, the $q$-profile is raised via generation of an interchange mode at the axis that adjusts the loop voltage through dynamo.
Steady state turbulence, which to some degree is always present in tokamaks
can co-exist with other modes and hence, not contradicting any of the previously mentioned works,
can explain why there is a stationary non-sawtooting state over resistive time scales.
In Fig.~\ref{fig:Lmode1}b we show how the safety
factor changes when various combinations of turbulent coefficients are acting together. The full line
represents the slightly reversed shear obtained from Ohmic and bootstrap current, while the upper two profiles (boxes and triangles) show $q_{saf}$ when anomalous resistivity is included. The rest (circles and crosses) are without anomalous resistivity, and do not significantly change the $q$-profile.
We conclude that the raising of the value of $q_{saf}$ is caused by
current reduction by the anomalous term at the axis, making this term
a possible candidate for suppression of sawteeth crashes. In Fig.~\ref{fig:Lmode1}c we show how the $q$-factor on the axis
changes with the level of turbulence $\tilde{b}$ for a $1.15 MA$ discharge with $q_0=0.983$ on the axis.
We note that certain level (threshold) of turbulence is required to raise the profile above $q_0=1$.
For $\tilde{b}<10^{-5}$ the value of $q_{saf}$ is diminishing slightly with $\tilde{b}$ before it starts to grow. This is due to the hyper-resistivity
which always flattens the surplus bootstrap current at the axis. For larger $\tilde{b}$ the effects of anomalous resistivity start to dominate and the safety factor is raised. From this discussion it becomes obvious that the turbulent threshold for
stabilization of the sawtooth is related to the thermodynamic profiles and the bootstrap current size and peak location.
\begin{figure}
\epsfxsize=0.7\linewidth
\centerline{\epsfbox{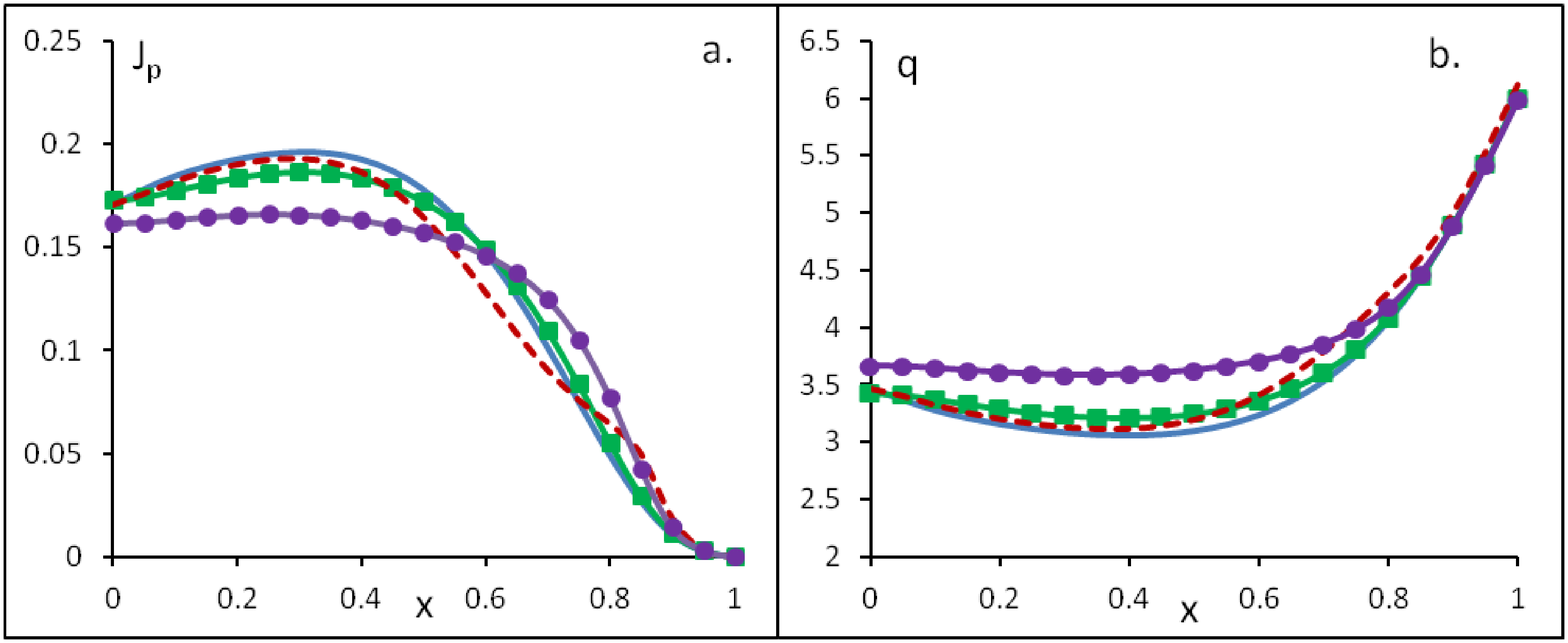}}
\caption{{\rm a}. Parallel current;  {\rm b}. Safety factor of the broad L-mode. Full line is for Ohmic+Bootstrap
current only, boxes- with axis turbulence, dashed line- with edge turbulence, circles- $O(10^{-4})$ axis turbulence. \label{fig:Lmode2}}
\end{figure}

For the broad L-mode given with boxes in Fig.\ref{fig:profile}b the
bootstrap current constitutes $19.8\%$ of the total current, and is radially wider compared to the one of the peaked mode, thus
giving the current profile represented with full line in Fig.\ref{fig:Lmode2}a. The shear is reversed over a wide region, which has some advantages concerning confinement and MHD stability (see next section for details). When the
turbulence is peaked on axis the final current after redistribution is given with boxes in Fig.\ref{fig:Lmode2}a,
while for turbulence peaked at the edge the current is significantly modified (Fig.~\ref{fig:Lmode2}a - dashed line) due to the alignment of the turbulence peak with strong temperature gradients.
The power is generated as ${\cal T}_{\rm \times} = 2.8\%$ and ${\cal T}_{\rm BS} = 20.3\%$, with
majority still coming from the inductive current.
The anomalous resistivity still reduces the total
current, but generates a bump near the edge which cannot be diffused towards the center. The total
current redistribution is about $5\%$, and it goes up to $16.8\%$ for $\tilde{b}=O(10^{-4})$ (line with circles), while for  $\tilde{b}=O(10^{-5})$ it is below $1\%$.

\subsection{Advanced-like scenario}
\label{advanced}

Generally speaking any scenario which has significantly
improved confinement and MHD stability over the standard H-mode can be referred to as advanced scenario.
These modes have broad current profile and larger
bootstrap current compared to the conventional (peaked) modes.
A broad current affects negatively the
confinement, since the confinement time grows with the current peaking \cite{Peeters00}.
Additionally,
MHD stability suffers since both the external kink mode \cite{Wesson78} and ballooning mode are more unstable for broader currents \cite{Peeters00}.
However, the presence of a transport barrier along with the large bootstrap current
in advanced scenarios create a reversed shear in a broad region, which makes the ballooning modes as well as the neoclassical
tearing mode stable. The kink modes can be stabilized, for relatively high $\beta_N$ using
a conducting wall \cite{Kessel94}.
\begin{figure}
\epsfxsize=0.7\linewidth
\centerline{\epsfbox{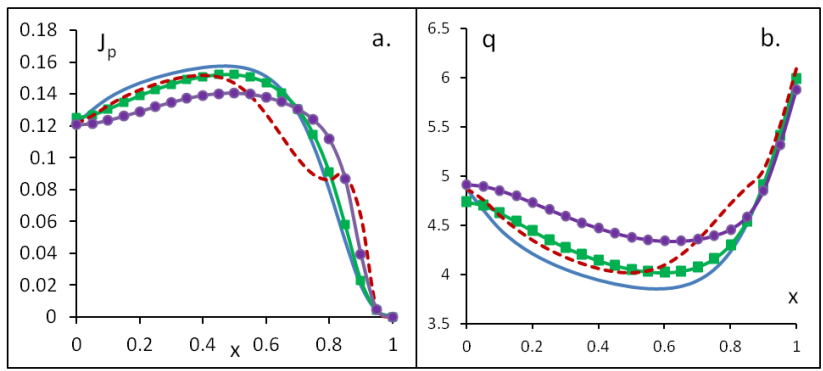}}
\caption{{\rm a}. Parallel current and {\rm b}. Safety factor of advanced scenario. Full line is for Ohmic+Bootstrap
current only, boxes- added axis turbulence, dashed line- edge turbulence, circles- $O(10^{-4})$ axis turbulence.\label{fig:advan}}
\end{figure}

In this section we will use the temperature profile given with circles in Fig.~\ref{fig:profile} to simulate an advanced scenario. This mode has a broad enough profile with a bootstrap current peaked near the edge and constituting $25\%$ of the total current. When the turbulence is peaked on the axis, there is negligible current generated from it. The profile remains broad, and becomes less hollow at the axis (Fig.\ref{fig:advan}a-boxes). More importantly the shear is still reversed over a wide region (Fig.\ref{fig:advan}b-boxes), making this in a way the ideal
scenario. The bootstrap current generates ${\cal T}_{\rm BS} = 24.9\%$ of the power, cross-resistivity ${\cal T}_{\rm \times} = 2.8\%$, while the anomalous and hyper-resistivity dissipate below $1\%$ of the power.
When speaking of anomalous resistivity the case of advanced scenario with edge turbulence (dashed lines in Fig.\ref{fig:advan}) deserves special attention. In this scenario the anomalous resistivity generates ${\cal T}_{\rm an} = 5.6\%$ of the power, which is similar to the cross-resistivity with ${\cal T}_{\rm \times} = 5.8\%$ and about $1/4$ of the bootstrap current with ${\cal T}_{\rm BS} = 23.8\%$. By turning on and off each turbulent coefficient, we conclude that the bump on the current near the edge is caused by the anomalous resistivity. This occurs due to the alignment of the region of negative anomalous resistivity with the temperature gradient and turbulence peak. This edge current does not diffuse towards the center, giving large redistribution of the current ($16.6\%$), and visibly changing the $q$-profile (dashed line in Fig.\ref{fig:advan}b). Cross-resistivity increases the total current by about $5\%$, but this effect is matched by the anomalous resistivity current reduction. Even for stronger turbulence $O(10^{-4})$ (line with circles) the total current is only increased by $2\%$. Since experimentally the exact current profile is difficult to measure especially near the axis, considering only the total parallel current most experiments would report negligible effects from turbulence.

\subsection{Fully bootstrapped tokamak}
\label{fully}

\begin{figure}
\epsfxsize=0.7\linewidth
\centerline{\epsfbox{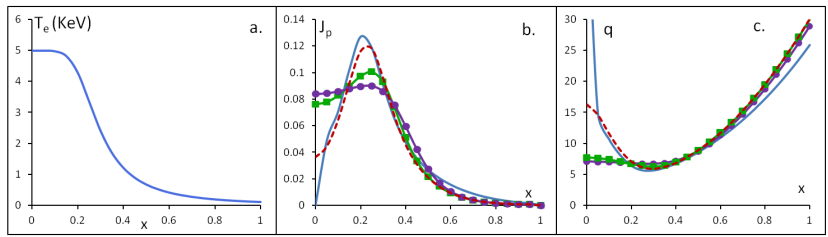}}
\caption{Fully bootstrapped tokamak; {\rm a}. Temperature profile; {\rm b}.
Current profile showing the bootstrap current (full line),  boxes- added axis turbulence, dashed line- edge turbulence,
circles- $O(10^{-4})$ axis turbulence; {\rm c}. Safety factor; \label{fig:fully}}
\end{figure}

These scenarios are usually related to internal transport barriers, where the current is generated by the strong pressure gradient, and the confinement is maintained by power generation by the bootstrap current through an internal loop. The alignment of the bootstrap current with the internal transport barrier is hence essential for the maintenance of the regime. In this section we use the temperature profile (in shape, not the absolute value) from a steady state, fully bootstrapped discharge in the TCV tokamak \cite{Coda08,Coda07}. The experimental data is fitted with the formula $T_e=0.1/\sqrt{x^{5.5} + 0.02^2}$ keV to obtain the temperature profile in Fig.\ref{fig:fully}a. For the ion temperature we use the same formula as the previous sections with $\gamma_T=2.0$, $\sigma_T=2.0$, and boundary values $T_{e,a}= 0.25 $ keV  and $T_{e,0}= 1.25$ keV. The density is the same as every simulation so far. With this equilibrium the current is generated $100\%$ via the bootstrap effect ($E_\parallel^0=0$). The full line in Fig.~\ref{fig:fully}b shows the bootstrap current as calculated by our model (compare to Fig.~4 of Ref.~\cite{Coda07}). The broadness of the profile is smaller compared to the advanced scenario, since the transport barrier is near the axis. Our code is steady-state, so it can't show how the internal current-temperature loop maintains the confinement. For a detailed description of how the steady state is achieved we refer the reader to Ref.~\cite{Coda08}.

In this scenario it's not unusual to have high level of perturbations around the axis, since the temperature gradient is strong in this area. The plots with boxes and dashed line in Fig.~\ref{fig:fully}b are for the turbulence profiles chosen so far using hyper-resistivity and cross-resistivity, while the line with circles is for stronger turbulence ($O(10^{-4})$). We note that hyper-resistivity is still effective in diffusing the current towards the center. The dissipation by hyper-resistivity is significant in this case (${\cal T}_{\rm H} = 9\%$) and the total current reduction is about $20\%$, while when cross-resistivity is added additional current is generated, but the total current is still smaller by $10\%$ than the initial bootstrap current. The cross-resistivity generates about ${\cal T}_{\rm \times}=18\%$ of the power, while the rest ${\cal T}_{\rm BS}=82\%$ comes from the bootstrap current. With the anomalous coefficient present, however, we could not find a steady state solution of the equation. While the hyper and cross-resistivity ``push'' the current toward the axis with some generation in the outer region by the cross-resistivity, the anomalous resistivity, which in this scenario is unusually high, significantly reduces the current at the axis, thus affecting the magnetic field and safety factor stability. Whether the presence of anomalous resistivity is really detrimental to the stability of the profiles is something that should be studied in coordination with a more complete transport code.

The other two coefficient do exactly what is necessary to deal with the hollow current profile, i.e. they diffuse the current towards the center and flatten the profile. This however results with change of the safety factor from strongly reversed in the region with transport barrier, to a safety factor which is only slightly reversed (Fig.~\ref{fig:fully}c) over a broad region. This means that due to turbulence some of the mentioned confinement advantages of the reversed shear are lost. It should be mentioned here that obtaining higher bootstrap fraction and higher confinement are two different goals of the tokamak development and today's tokamaks are not designed for optimal bootstrap effects, but for confinement and MHD stability. When the turbulence is peaked off-axis the profiles remain similar to those given in Fig.~\ref{fig:fully} and hence the same conclusions apply in that case too.
When the fluctuation level is of $O(10^{-5})$ the turbulence can still redistribute some current at the axis, however this effect is smaller and the current profile remains hollow. In this case additional effects like potato orbits should be taken into account.

\section{Summary and conclusions}
\label{summary}

Numerically solving a turbulent version of Ohm's law
Eq.~(\ref{ohmsteady1}) we have extended the study presented in Ref.~\cite{GattoOPPJ2011}
to include advanced plasma regimes, broad L-modes and
fully bootstrapped tokamaks, modeling the magnetic
turbulence with two different spectra - one peaked on-axis and one off-axis.
Three turbulent terms are present in the generalized Ohm's law: the hyper-resistive term, the anomalous
resistivity, and
the 'cross' resistive term, the latter being proportional to the derivative of the current density.
The anomalous resistivity is the largest coefficient, although all three of them are much smaller that
the neoclassical resistivity. Despite this, we find that the turbulent terms have an important impact on
the equilibrium current density and safety factor profiles, and thus should be retained in the Ohm's law.
The results of our study better elucidate the impact that each turbulent contribution has on shaping the
current and safety factor profiles, as well as on the power generation inside the plasma.
Aside from the magnitude of the turbulent perturbations, a key feature characterizing the various scenarios is the degree of alignment between the temperature gradient and the turbulence peak.

In L-modes and advanced scenarios, the well-known dissipative hyper-resistive term reduces the total current while diffusing the bootstrap current toward the plasma center. When both bootstrap current and turbulence profile are peaked in the outer region, the hyper-resistive diffusion is less efficient. The hyper-resistivity is most important in scenarios with high bootstrap current peaked near the axis, such as fully bootstrapped tokamak scenarios, where it plays a crucial role in maintaining a non-hollow current profile. The anomalous resistivity leads to a significant reduction of the current density in the central part of the plasma, and to a small increase in the outer region where it has negative value due to the combined effect of the thermodynamic and magnetic equilibrium profiles. Although usually dissipative, this term can generate power when there is an alignment of it's negative peak with the temperature gradient and the turbulent spectrum, which is the case in some advanced scenarios. The cross-resistive term, contrary to the anomalous resistive term, always amplifies the current by roughly $20-25\%$ of the bootstrap current, depending on the scenario. This term is most commonly generating power, except in rare cases when there is a strong current peak coinciding with the region of negative slope of the cross-resistivity (see Eq.~(\ref{power_balance})).

While the competing effects of the three turbulent coefficients on the integral parallel current inside the plasma cancel out in most cases, the redistribution of current especially around the axis causes the q-profile to change, which might result with a suppression of sawteeth crashes in some regimes. When the turbulence is weaker, i.e. $\tilde{b}\lesssim O(10^{-5})$ in the entire plasma region, its effects become less significant, even negligible. A very-broad temperature profile with a steep barrier near the edge has been used to study advanced-like scenarios. With on-axis turbulence, the equilibrium profiles are not different in a substantial way from the case with no turbulence. The presence of edge-localized turbulence, on the contrary, leads to a notable increase in power (due to anomalous and cross resistivity), while the increase in current is small. Additionally, turbulence produces an irregularity in the edge current profile, which the
hyper-resistive term is not able to smooth out. A fully bootstrapped scenario has been studied by adopting a temperature profile that models a steady-state, fully bootstrapped discharged in the TCV tokamak. This profile leads to a transport barrier which is located closer to the axis. Adopting the on-axis turbulence  model, as plausible with this kind of thermodynamic profiles, we find an effective inward diffusion of the current due mainly to the hyper-resistive term. The power balance Eq.~(\ref{power_balance}), shows that the power generated internally by the neoclassical and turbulent bootstrap terms can compensate for the dissipation by the remaining terms, thus eliminating the need for an external source.

The  theoretical  understanding  of  the  influence  of  turbulence  on  Ohm's  law  in
tokamaks is a very difficult task, requiring complex kinetic calculation. While we realize
that a reasonably accurate, and widely accepted theory is still lacking, we also note that
there are features that are common to most theories that consider extension of the
Ohm's law due to turbulent effects \cite{GattoOPPJ2011,Itoh88,Hinton04,McDevitt13}.
We mention the work of Ref.~\cite{Garbet14}, where it is estimated that turbulence can
generate a current up to $10\%$ of the neoclassical bootstrap, a potentiality built in our
model, too, as the numerical studies of the present work show. In the cited work, the
effective electric field (due to turbulence) was proportional to $k_\parallel R_0$, and thus a symmetry
breaking mechanism (e.g., a shear flow, or an inhomogeneous turbulent intensity) was
required in order to have a nonzero contribution.
In our current work we have not invoked a symmetry breaking mechanism,
so the integration in $k_\parallel$ over the normal modes of  magnetic
turbulence with $k_\parallel \ll k_\perp$ cancels out, and gives no contribution to the final
expression of the turbulent electric field. The overall effects in $k_\perp$,
which are non-zero, stem out from the self-consistency of our theory
(as shown in Ref.~\cite{Chavdarovski09}). It is a well documented fact that retaining the friction term
in the collision operator could lead, in same cases, to results quite different from the
corresponding results obtained with quasi-linear theory. We cite two examples: Ref.~\cite{Mynick92},
in which the energy exchange between species is evaluated using both quasi-linear and
self-consistent theory, and Ref.~\cite{Gatto2007}, which shows that self-consistency leads to particle
pinches.

One of the setbacks of this work is the use of circular flux surfaces and
classical instead of neoclassical resistivity, both approximations strictly
valid for cylindrical plasmas. Even with this simplified model we were able to
recover earlier results and give some new
insights into the effects of each turbulent coefficient on the current and magnetic profiles
in various scenarios.
The turbulent perturbation profile used here is fairly arbitrary and it could be
improved by reverse engineering, i.e. matching the current density experimental data with the
theoretical predictions. First step to expand this work will be to use turbulence profiles
generated from a more advanced numerical code.
We already mentioned that the current density itself affects the temperature and density
profile and with that
the general plasma stability. This is especially important when a
large fraction of the current
is due to bootstrap, in which case the mutual
interaction between the current and thermodynamic profiles must be taken into
account self-consistently. This has not been done here, and should be addressed
by coupling the Ohm's law to a transport code evolving the equilibrium
profiles. And last, in all computations we have set $j_\para=0$
at the plasma outer boundary. Thanks to turbulent diffusion, however,
a nonzero current at the plasma edge could potentially sustain
a significant fraction of the plasma current~\cite{Ono87}.

\vspace{0.50cm}
\noindent
{\bf Acknowledgment}

Useful discussions with C. Angioni, E. Poli, S. G\"unter and P.Lauber are kindly acknowledged. This work has been part of Max Planck-Princeton (MPPC)
Center for Plasma Physics study 2015-2017, and has been partially supported by ITER-CN project No. 2015GB110003.

\end{document}